\documentclass{ws-rv975x65}  
\usepackage{graphicx}

\begin{document}
\newcommand{\be}{\begin{equation}}
\newcommand{\ee}{\end{equation}}
\newcommand{\bea}{\begin{eqnarray}}
\newcommand{\eea}{\end{eqnarray}}
\newcommand{\ba}{\begin{array}}
\newcommand{\ea}{\end{array}}
\newcommand{\bc}{\begin{center}}
\newcommand{\ec}{\end{center}}
\renewcommand{\slash}{\!\!\!\!/\,}
\newcommand{\sslash}{\!\!\!/\,}
\newcommand{\Dslash}{D\hspace*{-0.23cm}/\,}
\newcommand{\dslash}{\partial\hspace*{-0.23cm}/\,}

\setcounter{chapter}{0}

\chapter{The CFL phase and $m_s$: An effective field theory 
approach}

\markboth{T.~Sch\"afer}{{The CFL phase and $m_s$}}

\author{Thomas Sch\"afer}

\address{Department of Physics\\
     North Carolina State University\\
     Raleigh, NC 27695}

\begin{abstract}
We study the phase diagram of dense quark matter with an 
emphasis on the role of the strange quark mass. Our approach 
is based on two effective field theories (EFTs). The first is 
an EFT that describes quark quasi-particles near the Fermi 
surface. This EFT is valid at energies small compared to the 
chemical potential. The second is an EFT for the Goldstone modes 
in the paired phase. We find that in response to a non-zero strange 
quark mass the CFL phase first undergoes a transition to a kaon 
condensed phase, and then to a gapless phase with a non-zero 
Goldstone boson current.
\end{abstract}

%%%%%%%%%%%%%%%%%%%%%%%%%%%%%%%%%%%%%%%%%%%%%%%%%%%%%%%%%%%%%%%%%%%%
\section{Introduction}     
\label{sec_intro} 
%%%%%%%%%%%%%%%%%%%%%%%%%%%%%%%%%%%%%%%%%%%%%%%%%%%%%%%%%%%%%%%%%%%%
 
  Searching for exotic states of matter at high baryon density 
and high temperature is one of the central efforts in nuclear
and particle physics. Calculations based on weak-coupling QCD 
indicate that the ground state of three flavor baryonic matter 
at very high density is the color-flavor-locked (CFL) phase 
\cite{Alford:1999mk,Schafer:1999fe,Evans:1999at}. The CFL phase
is characterized by a pair condensate
\be
\label{cfl}
 \langle \psi^a_i C\gamma_5 \psi^b_j\rangle  = 
  (\delta^a_i\delta^b_j-\delta^a_j\delta^b_i) \phi .
\ee
This condensate leads to a gap in the excitation spectrum 
of all fermions and completely screens the gluonic interaction. 
Both the chiral $SU(3)_L\times SU(3)_R$ and color $SU(3)$ 
symmetry are broken, but a vector-like $SU(3)$ flavor symmetry 
remains unbroken. 

  At baryon densities relevant to compact stars dis\-tor\-tions 
of the ideal CFL state due to quark masses cannot be neglected 
\cite{Alford:1999pa,Schafer:1999pb}. The most important effect of a 
non-zero strange quark mass is that the light and strange quark 
Fermi momenta will no longer be equal. When the mismatch is much 
smaller than the gap one finds for degenerate quarks, we expect 
that it has little consequence, since at this level the original 
particle and hole states near the Fermi surface are mixed up anyway. 
On the other hand, when the mismatch is much larger than the gap,
we expect that the ordering one finds in the symmetric system is 
disrupted, and that to a first approximation one can treat the 
light and heavy quark dynamics separately.

  This argument is qualitatively right, but the correct picture 
turns out to be much more complicated, and much more interesting. 
The phase diagram of cold dense strange quark matter contains 
phases with kaon and eta condensates, phases with non-zero currents, 
and crystalline states. Performing systematic calculations of the 
properties of these states is not straightforward, even if the 
coupling is weak. A standard set of tools are Dyson-Schwinger 
equations. This approach becomes quite involved once the strange quark
mass is taken into account, because there are many more gap parameters, 
and maintaining electric neutrality and color gauge invariance 
is difficult \cite{Alford:2002kj,Neumann:2002jm,Steiner:2002gx}. 
However, since chiral symmetry is broken in the CFL phase we know 
that the dependence on the quark masses is constrained by chiral 
symmetry. It is therefore natural to study the problem using 
effective field theories. In practice we will employ a two-step 
procedure. In the first step we match the microscopic theory, QCD, 
to an effective field theory of quasi-particles and holes in the 
vicinity of the Fermi surface. In the second step we match this 
theory to an effective chiral theory for the CFL phase.

%%%%%%%%%%%%%%%%%%%%%%%%%%%%%%%%%%%%%%%%%%%%%%%%%%%%%%%%%%%%%%%%%%%%%%%
\section{High density effective theory}
\subsection{Effective field theory near the Fermi surface}
\label{sec_hdet}
%%%%%%%%%%%%%%%%%%%%%%%%%%%%%%%%%%%%%%%%%%%%%%%%%%%%%%%%%%%%%%%%%%%%%%%

 The QCD Lagrangian in the presence of a chemical potential is given by
\be
\label{qcd}
 {\mathcal L} = \bar\psi \left( i\Dslash +\mu\gamma_0 -M \right)\psi
 -\frac{1}{4}G^a_{\mu\nu}G^a_{\mu\nu},
\ee
where $D_\mu=\partial_\mu+igA_\mu$ is the covariant derivative, $M$ 
is the mass matrix and $\mu$ is the baryon chemical potential. If the 
baryon density is very large perturbative QCD calculations can be 
simplified. The main observation is that the relevant degrees of 
freedom are particle and hole excitations in the vicinity of the 
Fermi surface. We shall describe these excitations in terms of the 
field $\psi_v(x)$, where $v$ is the Fermi velocity. At tree level, 
the quark field $\psi$ can be decomposed as $\psi=\psi_{v,+}+\psi_{v,-}$ 
where $\psi_{v,\pm}=P_{v,\pm}\psi$ with $P_{v,\pm}=\frac{1}{2}(1
\pm\vec{\alpha}\cdot\hat{v})\psi$. Note that $P_{v,\pm}$ is a projector
on states with positive/negative energy. To leading order in $1/\mu$ 
we can eliminate the field $\psi_-$ using its equation of motion. The 
lagrangian for the $\psi_+$ field is given by 
\cite{Hong:2000tn,Hong:2000ru,Beane:2000ms}
\be
\label{l_hdet}
{\cal L} = \psi_{v}^\dagger \left(
   iv\cdot D - \frac{D_\perp^2}{2\mu}
   -\frac{g\sigma_{\mu\nu}G_\perp^{\mu\nu}}{4\mu} \right) \psi_{v}
 -\frac{1}{4}G^a_{\mu\nu} G^a_{\mu\nu} + \ldots .
\ee
with $v_\mu=(1,\vec{v})$. Note that $v$ labels patches on the 
Fermi surface, and that the number of these patches grows
as $\mu^2$. The leading order $v\cdot D$ interaction does 
not connect quarks with different $v$, but soft gluons can 
be exchanged between quarks in different patches. In addition
to that, there are four, six, $\ldots$ fermion operators
that contain fermion fields with different velocity labels. 
These operators are constrained by the condition that the sum 
of the velocities has to be zero. 

%%%%%%%%%%%%%%%%%%%%%%%%%%%%%%%%%%%%%%%%%%%%%%%%%%%%%%%%%%%%%%%%%%%%
\begin{figure}[t]
\bc\includegraphics[width=7.5cm]{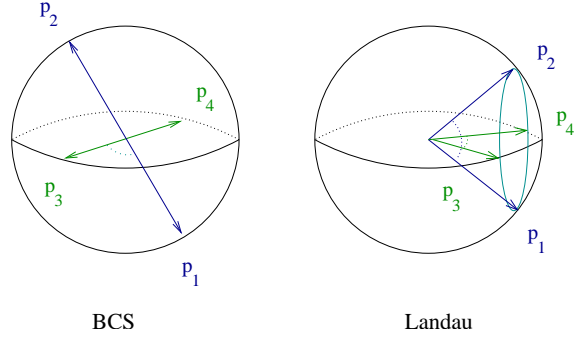}\ec
\caption{\label{fig_fskin}
Kinematics of four-fermion operators in the effective theory.}
\end{figure}
%%%%%%%%%%%%%%%%%%%%%%%%%%%%%%%%%%%%%%%%%%%%%%%%%%%%%%%%%%%%%%%%%%%%

 In the case of four-fermion operators there are two kinds of interactions 
that satisfy this constraint, see Fig.~\ref{fig_fskin}. The first possibility 
is that both the incoming and outgoing fermion momenta are back-to-back. 
This corresponds to the BCS interaction
\be
\label{c_bcs}
{\cal L}=  \frac{1}{\mu^2}\sum_{v',\Gamma,\Gamma'}
  V_l^{\Gamma\Gamma'} R_l^{\Gamma\Gamma'}(\vec{v}\cdot\vec{v}')
    \Big(\psi_{v} \Gamma \psi_{-v}\Big)
   \Big(\psi^\dagger_{v'}\Gamma'\psi^\dagger_{-v'}\Big),
\ee
where $\vec{v}\cdot\vec{v}'=\cos\theta$ is the scattering angle, 
$R_l^{\Gamma\Gamma'}(x)$ is a set of orthogonal 
polynomials, and $\Gamma,\Gamma'$ determine the color, flavor and spin 
structure. The second possibility is that the final momenta are 
equal to the initial momenta up to a rotation around the axis
defined by the sum of the incoming momenta. The relevant
four-fermion operator is
\be
\label{c_flp}
{\cal L}=  \frac{1}{\mu^2}\sum_{v',\Gamma,\Gamma'}
  F_l^{\Gamma\Gamma'} R_l^{\Gamma\Gamma'}(\vec{v}\cdot\vec{v}')
    \Big(\psi_{v} \Gamma \psi_{v'}\Big)
   \Big(\psi^\dagger_{\tilde{v}}\Gamma'\psi^\dagger_{\tilde{v}'}
  \Big).
\ee
In a system with short range interactions only the quantities
$F_l(0)$ are known as Fermi liquid parameters.

 The effective field theory expansion is complicated by the 
fact that the number of patches $N_v\sim \mu^2/\Lambda^2$ grows 
with the ratio of chemical potential $\mu$ over the EFT cutoff 
$\Lambda$. This implies that some higher order contributions that 
are suppressed by $1/\mu^2$ can be enhanced by powers of $N_v$. The 
natural solution to this problem is to sum the leading order diagrams 
in the large $N_v$ limit \cite{Schafer:2004zf}. In the gluon sector 
this corresponds to summing particle-hole loops in gluon $n$-point 
functions. There is a simple generating functional for these loop 
integrals which is known as the hard dense loop (HDL) effective action 
\cite{Braaten:1991gm}
\be
\label{S_hdl}
{\cal L}_{HDL} = -\frac{m^2}{2}\sum_v \,G^a_{\mu \alpha}
  \frac{v^\alpha v^\beta}{(v\cdot D)^2}G^b_{\mu\beta}.
\ee
In perturbation theory the dynamical gluon mass is given by
$m^2=N_f g^2\mu^2/(4\pi^2)$. In the fermion sector we have 
to consider four-fermion operators at leading order. Operators
with six or more fermion fields are suppressed. The effective 
lagrangian is given by
\bea
\label{l_udet}
 {\cal L}    &=& \psi_{v}^\dagger  \left(  i v\cdot D -
     \frac{D_\perp^2}{2\mu}   \right) \psi_{v}
    - \frac{1}{4}G^a_{\mu\nu}G^a_{\mu\nu} +   {\cal L}_{HDL} \\
& & \mbox{}+  \frac{1}{\mu^2}\sum_{v',\Gamma}\left[
  F_l^\Gamma R^\Gamma_l(x)(\psi_{v}\Gamma\psi_{v'})
             (\psi^\dagger_{v} \Gamma\psi^\dagger_{v'})
+ V_l^\Gamma R^\Gamma_l(x)(\psi_{v}\Gamma\psi_{-v})
             (\psi^\dagger_{v'} \Gamma\psi^\dagger_{-v'})
 \right] +\cdots .
 \nonumber
\eea

%%%%%%%%%%%%%%%%%%%%%%%%%%%%%%%%%%%%%%%%%%%%%%%%%%%%%%%%%%%%%%%%%%%%%%%
\subsection{Non-Fermi liquid effects}
\label{sec_nfl}
%%%%%%%%%%%%%%%%%%%%%%%%%%%%%%%%%%%%%%%%%%%%%%%%%%%%%%%%%%%%%%%%%%%%%%%

 In this Section we briefly study the effective field theory 
in the regime $\omega<m$ where $\omega$ is the excitation 
energy and $m$ is the effective gluon mass \cite{Schafer:2005mc}.
Since electric fields are screened the interaction at low energies 
is dominated by the exchange of magnetic gluons. The transverse 
gauge boson propagator is
\be
D_{ij}(k) = \frac{i(\delta_{ij}-\hat{k}_i\hat{k}_j)}
      {k_0^2-\vec{k}^2+i\frac{\pi}{2}m^2 \frac{k_0}{\vec{k}}} ,
\ee
where we have assumed that $|k_0|<|\vec{k}|$. We observe that the
propagator becomes large in the regime $|k_0|\sim |\vec{k}|^3/m^2$. 
If the energy is small, $|k_0|\ll m$, then the typical energy is 
much smaller than the typical momentum,
\be
\label{ld_kin}
 |\vec{k}| \sim (m^2 |k_0|)^{1/3} \gg |k_0| .
\ee
This implies that the gluon is very far off its energy shell
and not a propagating state. We can compute loop diagrams
containing quarks and transverse gluons by picking up the pole in 
the quark propagator, and then integrating over the cut in the gluon 
propagator using the kinematics dictated by equ.~(\ref{ld_kin}). 
In order for a quark to absorb the large momentum carried by the 
gluons and stay close to the Fermi surface this momentum has to 
be transverse to the momentum of the quark. This means that the 
term $k_\perp^2/(2\mu)$ in the quark propagator is relevant and 
has to be kept at leading order. Equation (\ref{ld_kin}) shows 
that $k_\perp^2/(2\mu)\gg k_0$ as $k_0\to 0$. This means that the 
pole of the quark propagator is governed by the condition $k_{||}
\sim k_\perp^2/(2\mu)$. We find
\be
\label{ld_reg}
 k_\perp \sim g^{2/3}\mu^{2/3}k_0^{1/3},\hspace{0.5cm}
 k_{||}  \sim g^{4/3}\mu^{1/3}k_0^{2/3}.
\ee
In the low energy regime propagators and vertices can be simplified
even further. The quark and gluon propagators are
\be
   S_{\alpha\beta}(p) = \frac{i\delta_{\alpha\beta}}
       {p_0-p_{||}-\frac{p_\perp^2}{2\mu}
              +i\epsilon {\it sgn}(p_0)},
\ee\be
   D_{ij}(k) = \frac{-i\delta_{ij}}
       {k_\perp^2-i\frac{\pi}{2}m^2\frac{k_0}{k_\perp}},
\ee
and the quark gluon vertex is $gv_i(\lambda^a/2)$. Higher order 
corrections can be found by expanding the quark and gluon propagators 
as well as the HDL vertices in powers of the small parameter $\epsilon
\equiv (k_0/m)$.

 We will refer to the regime in which all momenta, including external 
ones, satisfy the scaling relation (\ref{ld_reg}) as the Landau damping 
regime. The Landau damping regime is completely perturbative, i.e.~graphs 
with extra loops are always suppressed by extra powers of $\epsilon^{1/3}$. 
Every quark propagator scales as $\epsilon^{-2/3}$, gluon propagators scale 
as $\epsilon^{-2/3}$, and every loop integral gives $\epsilon^{7/3}$. The 
quark-gluon vertex scales as $\epsilon^0$ and the HDL three-gluon vertex 
scales as $\epsilon^{1/3}$. Using these results we can show that additional 
loops always increase the power of $\epsilon^{1/3}$ associated with the 
diagram.

 The effective theory describes a non-Fermi liquid. This is clear
from the appearance of fractional powers and logarithms in the 
low energy expansion. The simplest diagram which gives a logarithmic
term is the fermion self energy. We find 
\cite{Vanderheyden:1996bw,Manuel:2000mk,Brown:2000eh,Ipp:2003cj}
\be 
\label{sig_m}
 \Sigma(p) = \frac{g^2}{9\pi^2} 
    \left( p_0 \log\left(\frac{2^{5/2}m}{\pi|p_0|} \right) 
       + p_0 + i\frac{\pi}{2} p_0 \right)
       + O\left(\epsilon^{5/3}\right) .
\ee
The scale inside the logarithm is determined by matching the effective 
theory in the Landau damping regime to an effective theory that contains 
electric gluon exchanges. The $p_0\log(p_0)$ term leads to a vanishing
Fermi velocity as we approach the Fermi surface. 

 The effective theory can also be used to study other higher order 
corrections. We find, in particular, a QCD version of Migdal's 
theorem: In the Landau damping regime loop corrections to the 
quark-gluon vertex are suppressed by powers of $\epsilon^{1/3}$.

%%%%%%%%%%%%%%%%%%%%%%%%%%%%%%%%%%%%%%%%%%%%%%%%%%%%%%%%%%%%%%%%%%%%%%%
\subsection{Color superconductivity}
\label{sec_csc}
%%%%%%%%%%%%%%%%%%%%%%%%%%%%%%%%%%%%%%%%%%%%%%%%%%%%%%%%%%%%%%%%%%%%%%%

 It is well known that the particle-particle scattering amplitude in 
the BCS channel $q(\vec{p}\,)+q(-\vec{p}\,)\to q({\vec{p}}^{\,\prime})+q(
-{\vec{p}}^{\,\prime})$ is special. The total momentum of the pair 
vanishes and as a consequence loop corrections to the scattering 
amplitude are logarithmically divergent. This implies that all ladder 
diagrams have to be summed. Crossed ladders, vertex corrections, etc.~are 
perturbative and follow the scaling rules discussed in the previous section. 

 If the interaction in the particle-particle channel is attractive 
then the BCS singularity leads to the formation of a pair condensate 
and to a gap in the fermion spectrum. The gap can be computed by 
solving a Dyson-Schwinger equation for the anomalous (particle-particle)
self energy. In QCD the interaction is attractive in the color 
anti-triplet channel. The structure of the gap is simplest in the 
case of two flavors. In that case, there is a unique color 
anti-symmetric spin zero gap term of the form 
\be 
\label{2SC}
 \langle \psi^a_i C\gamma_5 \psi^b_j\rangle  \sim
\Delta\epsilon^{3ab}\epsilon^{ij}.
\ee
Here, $a,b$ labels color and $i,j$ flavor. The gap equation is given by
\cite{Son:1999uk,Schafer:1999jg,Pisarski:2000tv,Hong:2000fh}
\be
\label{eliash_mel}
\Delta(p_4) = \frac{g^2}{18\pi^2} \int dq_4
 \log\left(\frac{\Lambda_{BCS}}{|p_4-q_4|}\right)
    \frac{\Delta(q_4)}{\sqrt{q_4^2+\Delta(q_4)^2}},
\ee
where the scale $\Lambda_{BCS}=256\pi^4(2/N_f)^{5/2}g^{-5}\mu$ is 
again determined by electric gluon exchanges. The solution to the 
equation was found by Son \cite{Son:1999uk}. The value of the gap
on the Fermi surface $\Delta_0=\Delta(p_0\!=\!0)$ is 
\be
\label{gap_oge}
\Delta_0 \simeq 2\Lambda_{BCS}
   \exp\left(-\frac{\pi^2+4}{8}\right)
   \exp\left(-\frac{3\pi^2}{\sqrt{2}g}\right).
\ee
This result is correct up to $O(g)$ corrections to the pre-exponent.
In order to achieve this accuracy the $g^2\omega\log(\omega)$ term 
in the normal self energy, equ.~(\ref{sig_m}), has to be included
in the gap equation \cite{Brown:1999aq,Wang:2001aq,Schafer:2003jn}. 
The condensation energy is given by
\be
\label{eps_2sc}
\epsilon = -N_d \Delta_0^2\left(\frac{\mu^2}{4\pi^2}\right),
\ee
where $N_d=4$ is the number of condensed species. 

 The situation is slightly more complicated in QCD with $N_f=3$ 
massless flavors. One possibility is to embed the $N_f=2$ order 
parameter into $N_f=3$ QCD. This option is usually called the 
2SC phase. The energetically preferred phase is the CFL phase
described by the order parameter given in equ.~(\ref{cfl}). In 
the CFL phase there are eight fermions with gap $\Delta_{CFL}$ 
and one fermion with gap $2\Delta_{CFL}$. The value of the gap
in the CFL phase is smaller than the one in the 2SC phase by 
a factor\cite{Schafer:1999fe} $2^{-1/3}$. The condensation 
energy in the two phases is given by
\be 
\epsilon =-\Delta_0^2\left(\frac{\mu^2}{4\pi^2}\right)
\left\{ \begin{array}{cl} 4 & {\rm 2SC}\\ 
   \; 12\cdot 2^{-2/3} \; & {\rm CFL} \end{array}\right. 
\ee
and the CFL phase is preferred by a factor $(27/4)^{1/3}\simeq 1.89$.

%%%%%%%%%%%%%%%%%%%%%%%%%%%%%%%%%%%%%%%%%%%%%%%%%%%%%%%%%%%%%%%%%%%%%%%
\subsection{Mass terms}
\label{sec_mhdet}
%%%%%%%%%%%%%%%%%%%%%%%%%%%%%%%%%%%%%%%%%%%%%%%%%%%%%%%%%%%%%%%%%%%%%%%

 Mass terms modify the parameters in the effective lagrangian.
These parameters include the Fermi velocity, the effective 
chemical potential, the screening mass, the BCS terms and the
Landau parameters. At tree level the correction to the Fermi
velocity and the chemical potential are given by
\be 
\label{vf_mu}
 v_F= 1-\frac{m^2}{2p_F^2},\hspace{1cm}
\delta \mu =-\frac{m^2}{2p_F}.
\ee
The shift in the Fermi velocity also affects the coupling 
$gv_F$ of a magnetic gluon to quarks. It is important to note 
that at leading order in $g$ this the only mass correction
to the coupling. This is not entirely obvious, as one can 
imagine a process in which the quark emits a gluon, makes 
a transition to a virtual high energy state, and then couples
back to a low energy state by a mass insertion. This process
would give an $O(m/\mu)$ correction to $g$, but it vanishes
in the forward direction \cite{Schafer:2001za}. 

%%%%%%%%%%%%%%%%%%%%%%%%%%%%%%%%%%%%%%%%%%%%%%%%%%%%%%%%%%%%%%%%%%%%
\begin{figure}[t]
\bc\includegraphics[width=8.25cm]{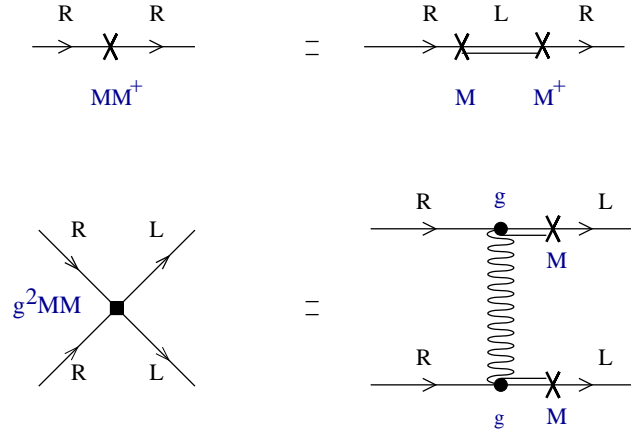}\ec
\caption{\label{fig_hdet_m}
Mass terms in the high density effective theory. The first 
diagram shows a $O(MM^\dagger)$ term that arises from integrating 
out the $\psi_-$ field in the QCD lagrangian. The second
diagram shows a $O(M^2)$ four-fermion operator which arises from 
integrating out $\psi_-$ and hard gluon exchanges.}
\end{figure}
%%%%%%%%%%%%%%%%%%%%%%%%%%%%%%%%%%%%%%%%%%%%%%%%%%%%%%%%%%%%%%%%%%%%

  Quark masses modify quark-quark scattering amplitudes and 
the corresponding Landau and BCS type four-fermion operators. 
Consider quark-quark scattering in the forward direction, 
$v+v'\to v+v'$. At tree level in QCD this process receives
contribution from the direct and exchange graph. In the 
effective theory the direct term is reproduced by the collinear 
interaction while the exchange terms has to be matched against 
a four-fermion operator. The spin-color-flavor symmetric 
part of the exchange amplitude is given by 
\be
\label{m_fl}
{\cal M}(v,v';v,v') =  \frac{C_F}{4N_c N_f}\frac{g^2}{p_F^2}
  \left\{ 1-\frac{m^2}{p_F^2}\frac{x}{1-x} \right\}
\ee
where $C_F=(N_c^2-1)/(2N_c)$ and $x=\hat{v}\cdot\hat{v}'$ is 
the scattering angle. We observe that the amplitude is independent 
of $x$ in the limit $m\to 0$. Mass corrections are singular as 
$x\to 1$. The means that the Landau coefficients $F_l$ contain
logarithms of the cutoff. We note that there is one linear combination 
of Landau coefficients, $F_0-F_1/3$, which is cutoff independent. 

 Equations (\ref{vf_mu}-\ref{m_fl}) are valid for $N_f\geq 1$ degenerate 
flavors. Spin and color anti-symmetric BCS amplitudes require at least 
two different flavors. Consider BCS scattering $v+(-v)\to v'+(-v')$
in the helicity flip channel $L+L\to R+R$. The color-anti-triplet 
amplitude is given by
\be 
\label{m_bcs}
{\cal M}(v,-v;v',-v') = \frac{C_A}{4}\frac{g^2}{p_F^2}
  \frac{m_1m_2}{p_F^2}.
\ee 
where $m_1$ and $m_2$ are the masses of the two quarks and $C_A=(N_c+1)/
(2N_c)$. We observe that the scattering amplitude is independent of the 
scattering angle. This means that at leading order in $g$ and $m$ only 
the s-wave potential $V_0$ is non-zero. 

 In order to match Green functions in the high density effective 
theory to an effective chiral theory of the CFL phase we need to 
generalize our results to a complex mass matrix of the form ${\cal L}=  
-\bar\psi_L M\psi_R - \bar\psi_R M^\dagger \psi_L $, see
Fig.~\ref{fig_hdet_m}. The $\delta\mu$ term is 
\be 
\label{m_kin}
{\mathcal L} = -\frac{1}{2p_F} \left( \psi_{L+}^\dagger MM^\dagger \psi_{L+}
 + \psi_{R+}^\dagger M^\dagger M\psi_{R+} \right).
\ee
and the four-fermion operator in the BCS channel is
\bea
\label{hdet_m}
 {\mathcal L} &=& \frac{g^2}{64p_F^4}
        ({\psi^A_L}^\dagger C{\psi^B_L}^\dagger)
        (\psi^C_R C \psi^D_R) 
   \Big\{ {\rm Tr} \left[ 
      \lambda^A M(\lambda^D)^T \lambda^B M (\lambda^C)^T\right] \\
 & & \hspace{4.2cm}\mbox{}
   -\frac{1}{3} {\rm Tr} \left[
    \lambda^A M(\lambda^D)^T \right]
    {\rm Tr} \left[
    \lambda^B M (\lambda^C)^T\right] \Big\}. \nonumber 
\eea
Here, we have introduced the CFL eigenstates $\psi^A$ defined by 
$\psi^a_i=\psi^A (\lambda^A)_{ai} /\sqrt{2}$, $A=0,\ldots,8$.

%%%%%%%%%%%%%%%%%%%%%%%%%%%%%%%%%%%%%%%%%%%%%%%%%%%%%%%%%%%%%%%%%
\subsection{Normal quark matter}
\label{sec_nqm}
%%%%%%%%%%%%%%%%%%%%%%%%%%%%%%%%%%%%%%%%%%%%%%%%%%%%%%%%%%%%%%%%%

 Before we consider quark mass effects in the superconducting 
phase we would like to review mass corrections in the normal 
phase of quark matter. We will consider three flavor quark 
matter with massless up and down quarks and massive strange 
quarks. There are some simple mass effects that we can 
directly deduce from equ.~(\ref{vf_mu}). The strange quark 
chemical potential is shifted by $\delta\mu_s=m_s^2/(2p_F)$
with respect to the strange quark Fermi momentum, and the 
strange quark Fermi velocity is smaller than one. 

 It would seem that perturbative corrections to these results
are sensitive to momenta far away from the Fermi surface and
cannot be computed in the framework of an effective field
theory for modes near $p_F$. Landau showed, however, that 
Galilei invariance leads to a relation between the chemical 
potential and the Fermi momentum that only involves the 
interaction on the Fermi surface. This relation was generalized 
to relativistic systems by Baym and Chin \cite{Baym:1975va}. 
They showed that 
\be   
\label{mu-pf}
\mu d\mu = \left[ p_F + \frac{p_F^2}{\pi^2} 
   \left( F_0^s - \frac{1}{3}F_1^s \right) \mu \right] dp_F,
\ee
where $F_{0,1}^s$ are color-flavor-spin symmetric Landau 
coefficients. We observe that this relation involves precisely 
the combination of Landau coefficients that does not depend on 
the EFT cutoff. This implies that we can integrate equ.~(\ref{mu-pf})
as in ordinary Landau Fermi Liquid theory. The result can be 
used to determine to thermodynamic potential to $O(g^2)$. We 
get \cite{Freedman:1976ub,Baluni:1977ms}
\bea
\Omega &=& - \frac{N_c}{12 \pi^2}
  \Bigg[\mu u(\mu^2-\frac{5}{2}m^2)+
  \frac{3}{2}m^4\ln{\left(\frac{\mu+u}{m}\right)}
  \Bigg] \nonumber \\
 & & \mbox{}+\frac{\alpha_s (N_c^2-1)}{16\pi^3}
  \Bigg[3 \left(m^2 \ln\frac{\mu+u}{m}-\mu u
      \right)^2-2 u^4 \Bigg] .
\eea
with $u=\sqrt{\mu^2-m^2}$. This result corresponds to a 
momentum space space subtraction scheme. The thermodynamic
potential in the $\overline{MS}$ scheme was computed by Fraga
and Romatschke \cite{Fraga:2004gz}. Perturbative corrections reduce 
the pressure of a quark gas. The same is true for mass corrections
to the pressure in the non-interacting system. The $O(\alpha_s m^2)$ 
term, however, increases the pressure. This is related to the 
fact that the exchange energy in a degenerate quark gas changes
sign in going from the non-relativistic limit, dominated by the 
Coulomb interaction, to the relativistic system in which 
magnetic interactions are more important. 

 In applications to neutron stars we have to include weak interactions 
and enforce electric charge neutrality. Weak interactions can convert 
strange quarks into up quarks, electrons and neutrinos. We will assume
that neutrinos can leave the system. Under these assumptions the system 
is characterized by a baryon chemical potential $\mu$ and an electron 
chemical potential $\mu_e$. The electron chemical potential is fixed 
by the condition of electric charge neutrality, $(\partial \Omega) 
/(\partial\mu_e)=0$. 

To leading order in $m_s^2/p_F^2$ the electron chemical potential 
is given by
\be
\label{mue}
\mu_e \simeq \frac{m_s^2}{4p_F}\left(
 1 -\frac{4\alpha_s}{\pi} \log\left(\frac{2p_F}{m_s}\right)\right).
\ee
At tree level we find $\mu_e=\mu_s/2$ and the Fermi surfaces
are split symmetrically 
\be 
p_F^d=p_F^u+\mu_s/2, \hspace{0.5cm}
p_F^s=p_F^u-\mu_s/2.
\ee
Perturbative corrections reduce the magnitude of $\mu_e$. In fact, 
since the $O(\alpha_s)$ term is enhanced by a large logarithm 
$\log(p_F/m)$, the electron chemical potential can become negative. 
This result is probably not reliable. In particular, the $O(\alpha_s)$ 
term is modified if the $\overline{MS}$ mass is used. 

%%%%%%%%%%%%%%%%%%%%%%%%%%%%%%%%%%%%%%%%%%%%%%%%%%%%%%%%%%%%%%%%%
\section{Chiral theory of the CFL phase}
\label{sec_CFLchi}
%%%%%%%%%%%%%%%%%%%%%%%%%%%%%%%%%%%%%%%%%%%%%%%%%%%%%%%%%%%%%%%%%

%%%%%%%%%%%%%%%%%%%%%%%%%%%%%%%%%%%%%%%%%%%%%%%%%%%%%%%%%%%%%%%%%
\subsection{Introduction}
\label{sec_CFLchi_intro}
%%%%%%%%%%%%%%%%%%%%%%%%%%%%%%%%%%%%%%%%%%%%%%%%%%%%%%%%%%%%%%%%%

 The main topic of this review is the effect of the strange quark
mass on the CFL phase. In principle this problem can be studied
by solving a gap equations which includes mass corrections to 
the chemical potential, the Fermi velocity, and the quark-quark
interaction. In full QCD this has not been attempted yet, but 
there are many calculations of this type that are based on 
effective four-fermion models of QCD. We will compare our results 
with some of these calculations in Sect.~\ref{sec_mic}.

 There are several difficulties with microscopic calculations
of this kind. The first is that they typically require an ansatz
for the gap parameter. We will see that both flavor symmetry 
and isospin are broken, and that the gaps for the left and 
right handed fermions acquire a relative phase. This means
that the number of independent gap parameters is quite large. 
The second difficulty is that while the ideal CFL state is
automatically neutral with respect to all charges, this is 
no longer the case once the state is perturbed by a finite 
strange quark mass. This means that we have to introduce flavor 
chemical potentials and gluonic background fields. Finally, 
we know that chiral symmetry is broken in the CFL phase and
this implies that physical observables are non-analytic in 
the quark mass.

 These non-analyticities are related to Goldstone bosons. This 
suggests that we should study the effects of a non-zero strange 
quark mass using an effective field theory for the Goldstone
modes. Indeed, it is well known that chiral symmetry places
important constraints on the mass dependence of QCD observables, 
and that these constraints are most easily implemented by using
an effective lagrangian.

%%%%%%%%%%%%%%%%%%%%%%%%%%%%%%%%%%%%%%%%%%%%%%%%%%%%%%%%%%%%%%%%%
\subsection{Chiral effective field theory}
\label{sec_CFLeft}
%%%%%%%%%%%%%%%%%%%%%%%%%%%%%%%%%%%%%%%%%%%%%%%%%%%%%%%%%%%%%%%%%

 For excitation energies smaller than the gap the only relevant 
degrees of freedom are the Goldstone modes associated with the 
breaking of chiral symmetry and baryon number. Since the pattern 
of chiral symmetry breaking is identical to the one at $T=\mu=0$ 
the effective lagrangian has the same structure as chiral perturbation 
theory. The main difference is that Lorentz-invariance is broken 
and only rotational invariance is a good symmetry. The effective 
lagrangian for the Goldstone modes is given by \cite{Casalbuoni:1999wu}
\bea
\label{l_cheft}
{\mathcal L}_{eff} &=& \frac{f_\pi^2}{4} {\rm Tr}\left[
 \nabla_0\Sigma\nabla_0\Sigma^\dagger - v_\pi^2
 \partial_i\Sigma\partial_i\Sigma^\dagger \right] 
 +\left[ B {\rm Tr}(M\Sigma^\dagger) + h.c. \right] 
    \nonumber \\ 
 & & \hspace*{0cm}\mbox{} 
     +\left[ A_1{\rm Tr}(M\Sigma^\dagger)
                        {\rm Tr} (M\Sigma^\dagger) 
     + A_2{\rm Tr}(M\Sigma^\dagger M\Sigma^\dagger) \right.
 \nonumber \\[0.1cm] 
  & &   \hspace*{0.5cm}\mbox{}\left. 
     + A_3{\rm Tr}(M\Sigma^\dagger){\rm Tr} (M^\dagger\Sigma)
         + h.c. \right]+\ldots . 
\eea
Here $\Sigma=\exp(i\phi^a\lambda^a/f_\pi)$ is the chiral field,
$f_\pi$ is the pion decay constant and $M$ is a complex mass
matrix. The chiral field and the mass matrix transform as
$\Sigma\to L\Sigma R^\dagger$ and  $M\to LMR^\dagger$ under 
chiral transformations $(L,R)\in SU(3)_L\times SU(3)_R$. We 
have suppressed the singlet fields associated with the breaking 
of the exact $U(1)_V$ and approximate $U(1)_A$ symmetries. 

%%%%%%%%%%%%%%%%%%%%%%%%%%%%%%%%%%%%%%%%%%%%%%%%%%%%%%%%%%%%%%%%%%%%
\begin{figure}[t]
\bc\includegraphics[width=10cm]{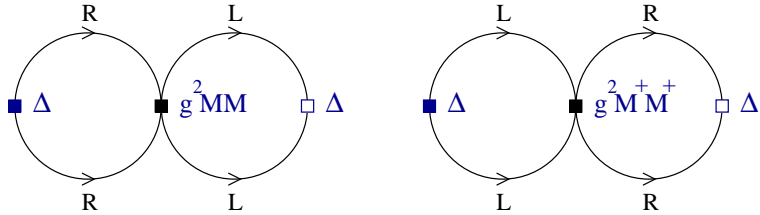}\ec
\caption{\label{fig_4fvac}
Contribution of the $O(M^2)$ BCS four-fermion operator to the 
condensation energy in the CFL phase .}
\end{figure}
%%%%%%%%%%%%%%%%%%%%%%%%%%%%%%%%%%%%%%%%%%%%%%%%%%%%%%%%%%%%%%%%%%%%

 At low density the coefficients $f_\pi$, $B,A_i,\ldots$ are 
non-perturbative quantities that have to extracted from 
experiment or measured on the lattice. At large density, on
the other hand, the chiral coefficients can be calculated in 
perturbative QCD. For the derivative terms this is most 
easily done by matching the two-point functions of flavor 
currents between QCD and the chiral theory. The mass terms
can be computed by matching the mass dependence of the vacuum
energy. At leading order in $\alpha_s$ the Goldstone boson decay 
constant and velocity are \cite{Son:1999cm} 
\be
\label{cfl_fpi}
f_\pi^2 = \frac{21-8\log(2)}{18} 
  \left(\frac{p_F^2}{2\pi^2} \right), 
\hspace{0.5cm} v_\pi^2=\frac{1}{3}.
\ee
Mass terms are determined by the operators studied in 
Sect.~\ref{sec_mhdet}. We observe that both equ.~(\ref{m_kin})
and (\ref{hdet_m}) are quadratic in $M$. This implies that $B=0$
in perturbative QCD. $B$ receives non-perturbative contributions
from instantons, but these effects are small if the density is
large, see Sect.~\ref{sec_inst}.

  We observe that $X_L=MM^\dagger/(2p_F)$ and $X_R=M^\dagger M/
(2p_F)$ in equ.~(\ref{m_kin}) act as effective chemical potentials 
for left and right-handed fermions, respectively. Formally, the 
effective lagrangian has an $SU(3)_L\times SU(3)_R$ gauge 
symmetry under which $X_{L,R}$ transform as the temporal components
of non-abelian gauge fields. We can implement this approximate gauge 
symmetry in the CFL chiral theory by promoting time derivatives
to covariant derivatives \cite{Bedaque:2001je}, 
\be
\label{mueff}
 \nabla_0\Sigma = \partial_0 \Sigma 
 + i \left(\frac{M M^\dagger}{2p_F}\right)\Sigma
 - i \Sigma\left(\frac{ M^\dagger M}{2p_F}\right) .
\ee
The BCS four-fermion operator in equ.~(\ref{hdet_m}) contributes to 
to the condensation energy in the CFL phase, see Fig.~\ref{fig_4fvac}.
We find \cite{Son:1999cm,Schafer:2001za}
\be
\label{E_MM}
\Delta {\cal E} = -\frac{3\Delta^2}{4\pi^2}
 \left\{  \Big( {\rm Tr}[M]\Big)^2 -{\rm Tr}\Big[ M^2\Big]
   \right\}
 + \Big(M\leftrightarrow M^\dagger \Big).
\ee
This term can be matched against the $A_i$ terms in the effective
lagrangian. The result is \cite{Son:1999cm,Schafer:2001za}
\be
 A_1= -A_2 = \frac{3\Delta^2}{4\pi^2}, 
\hspace{1cm} A_3 = 0.
\ee
The vacuum energy also receives contributions from Landau-type four-fermion 
operators, but these terms are proportional to ${\rm Tr}[MM^\dagger]$
and do not depend on the chiral field $\Sigma$.

 We can now summarize the structure of the chiral expansion in the
CFL phase. The effective lagrangian has the form 
\be
{\mathcal L}\sim f_\pi^2\Delta^2 \left(\frac{\partial_0}{\Delta}\right)^k
 \left(\frac{\vec{\partial}}{\Delta}\right)^l
 \left(\frac{MM^\dagger}{p_F\Delta}\right)^m
 \left(\frac{MM}{p_F^2}\right)^n
  \big(\Sigma\big)^o\big(\Sigma^\dagger\big)^p.
\ee
Loop graphs in the effective theory are suppressed by powers of 
$\partial/(4\pi f_\pi)$. Since the pion decay constant scales as 
$f_\pi\sim p_F$ Goldstone boson loops are suppressed compared to 
higher order contact terms. We also note that the quark mass expansion 
is controlled by $m^2/(p_F\Delta)$. This is means that the chiral
expansion breaks down if $m^2\sim p_F\Delta$. This is the same
scale at which BCS calculations find a transition from the CFL 
phase to a less symmetric state. 

%%%%%%%%%%%%%%%%%%%%%%%%%%%%%%%%%%%%%%%%%%%%%%%%%%%%%%%%%%%%%%%%%%%%
\begin{figure}[t]
\bc\includegraphics[width=7cm]{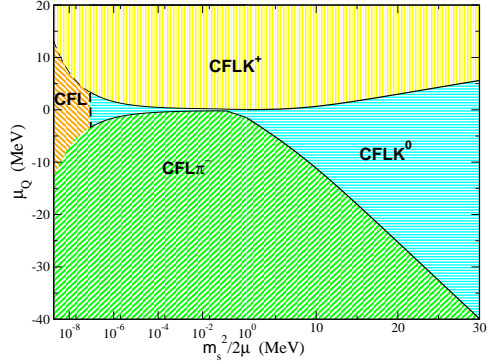}\ec
\caption{\label{fig_kcond}
Phase structure of CFL matter as a function of the effective 
chemical potential $\mu_s=m_s^2/(2p_F)$ and the lepton chemical 
potential $\mu_Q$, from Kaplan \& Reddy (2001). A typical value
of $\mu_s$ in a neutron star is 10 MeV. }
\end{figure}
%%%%%%%%%%%%%%%%%%%%%%%%%%%%%%%%%%%%%%%%%%%%%%%%%%%%%%%%%%%%%%%%%%%%

%%%%%%%%%%%%%%%%%%%%%%%%%%%%%%%%%%%%%%%%%%%%%%%%%%%%%%%%%%%%%%%%%%%%
\subsection{Kaon condensation}
\label{sec_kcond}
%%%%%%%%%%%%%%%%%%%%%%%%%%%%%%%%%%%%%%%%%%%%%%%%%%%%%%%%%%%%%%%%%%%%

 Using the chiral effective lagrangian we can now determine 
the dependence of the order parameter on the quark masses. We will 
focus on the physically relevant case $m_s>m_u=m_d$. Because
the main expansion parameter is $m_s^2/(p_F\Delta)$ increasing 
the quark mass is roughly equivalent to lowering the density. 
The effective potential for the order parameter is 
\be
\label{v_eff}
V_{eff} = \frac{f_\pi^2}{4} {\rm Tr}\left[
 2X_L\Sigma X_R\Sigma^\dagger -X_L^2-X_R^2\right] 
     - A_1\left[ \left({\rm Tr}(M\Sigma^\dagger)\right)^2 
     - {\rm Tr}\left((M\Sigma^\dagger)^2\right) \right].
\ee
The first term contains the effective chemical potential 
$\mu_s=m_s^2/(2p_F)$ and favors states with a deficit of 
strange quarks (with strangeness $S=-1$). The second term favors 
the neutral ground state $\Sigma=1$. The lightest excitation
with positive strangeness is the $K^0$ meson. We therefore
consider the ansatz $\Sigma = \exp(i\alpha\lambda_4)$ which
allows the order parameter to rotate in the $K^0$ direction. 
The vacuum energy is 
\be 
\label{k0+_V}
 V(\alpha) = -f_\pi^2 \left( \frac{1}{2}\left(\frac{m_s^2-m^2}{2p_F}
   \right)^2\sin(\alpha)^2 + (m_{K}^0)^2(\cos(\alpha)-1)
   \right),
\ee
where $(m_K^0)^2= (4A_1/f_\pi^2)m(m+m_s)$. Minimizing the vacuum 
energy we obtain 
\be 
\cos(\alpha)= \left\{ \begin{array}{cl}
 1 & \mu_s<m_K^0 \\
\;\frac{(m_K^0)^2}{\mu_s^2}\; & \mu_s >m_K^0\\
\end{array}\right.
\ee
The hypercharge density is 
\be 
n_Y = f_\pi^2 \mu_s \left( 1- \frac{(m_K^0)^4}{\mu_s^4}\right).
\ee
This result has the same structure as the charge density of a 
weakly interacting Bose condensate. Using the perturbative result 
for $A_1$ we can get an estimate of the critical strange quark mass.
We find  
\be
\label{ms_crit}
 m_s (crit)= 3.03\cdot  m_d^{1/3}\Delta^{2/3},
\ee
from which we obtain $m_s(crit)\simeq 70$ MeV for $\Delta\simeq 50$ 
MeV. This result suggests that strange quark matter at densities that 
can be achieved in neutron stars is kaon condensed. We also note that 
the difference in condensation energy between the CFL phase 
and the kaon condensed state is not necessarily small. For 
$\mu_s\to \Delta$ we have $\sin(\alpha)\to 1$ and $V(\alpha)\to 
f_\pi^2\Delta^2/2$. Since $f_\pi^2$ is of order $\mu^2/(2\pi^2)$
this is comparable to the condensation energy in the CFL phase. 

%%%%%%%%%%%%%%%%%%%%%%%%%%%%%%%%%%%%%%%%%%%%%%%%%%%%%%%%%%%%%%%%%%%%
\begin{figure}[t]
\bc \includegraphics[width=7cm]{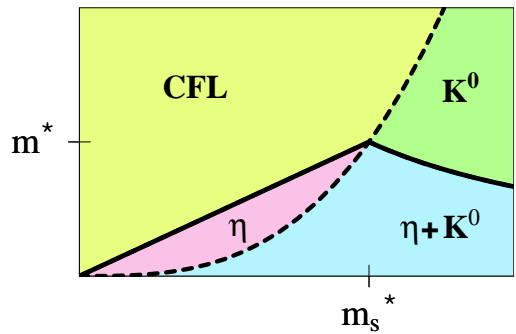}\ec
\caption{\label{fig_ph_ms_mu}
Phase structure of CFL matter as a function of the light quark mass
$m$ and the strange quark mass $m_s$, from Kryjevski, Kaplan \&
Sch\"afer (2005).}
\end{figure}
%%%%%%%%%%%%%%%%%%%%%%%%%%%%%%%%%%%%%%%%%%%%%%%%%%%%%%%%%%%%%%%%%%%%

  The strange quark mass breaks the $SU(3)$ flavor symmetry to 
$SU(2)_I\times U(1)_Y$. In the kaon condensed phase this symmetry 
is spontaneously broken to $U(1)_Q$. If isospin is an exact symmetry 
there are two exactly massless Goldstone modes \cite{Schafer:2001bq},
the $K^0$ and the $K^+$. Isospin breaking leads to a small mass for 
the $K^+$. The phase structure as a function of the strange quark mass 
and non-zero lepton chemical potentials was studied by Kaplan and Reddy 
\cite{Kaplan:2001qk}, see Fig.~\ref{fig_kcond}. We observe that if the 
lepton chemical potential is non-zero charged kaon and pion condensates 
are also possible. 

%%%%%%%%%%%%%%%%%%%%%%%%%%%%%%%%%%%%%%%%%%%%%%%%%%%%%%%%%%%%%%%%%%
\subsection{Eta meson condensation}
\label{sec_eta}
%%%%%%%%%%%%%%%%%%%%%%%%%%%%%%%%%%%%%%%%%%%%%%%%%%%%%%%%%%%%%%%%%%

 The CFL phase also contains a very light $S=0$ mode which can 
potentially become unstable. This mode is a linear combination 
of the $\eta$ and $\eta'$ and its mass is proportional to $m_u m_d$. 
Because this mode has zero strangeness it is not affected by the 
$\mu_s$ term in the effective potential. However, since $m_u,m_d
\ll m_s$ this state is sensitive to perturbative $\alpha_s m_s^2$ 
corrections. The relevant contribution to the effective lagrangian 
is \cite{Kryjevski:2004cw}
\be 
\delta{\cal L} = - \delta A \, 
  \left[({\rm Tr} M \Sigma^\dagger)^2 + 
         {\rm Tr} (M\Sigma^\dagger)^2 \right] 
  + {\rm h.c.}\ ,
\ee
with 
\be 
\delta A   = \frac{3}{8\pi^2} \Delta_6^2, \hspace{1cm}
\Delta_6^2 = \alpha_s\frac{( \ln 2)^2\,}{162\pi}\Delta^2\ .
\label{dsix}
\ee
Here, $\Delta_6$ is a color-flavor symmetric gap parameter which 
is generated by perturbative corrections to the dominant, color-flavor
anti-symmetric gap \cite{Schafer:1999fe}. Since the one-gluon exchange 
interaction in the color-symmetric channel is repulsive the $O(\alpha_s 
m_s^2)$ contribution to the mass of the $\eta-\eta'$ mode tends to 
cancel the $O(m_um_d)$ term. When the two terms become equal the eta 
condenses. The resulting phase diagram is shown in Fig.~\ref{fig_ph_ms_mu}.
The precise value of the tetra-critical point $(m^*,m_s^*)$ depends
sensitively on the value of the coupling constant. At very high 
density $m^*$ is extremely small, but at moderate density $m^*$ 
can become as large as 5 MeV, comparable to the physical 
values of the up and down quark mass. 

%%%%%%%%%%%%%%%%%%%%%%%%%%%%%%%%%%%%%%%%%%%%%%%%%%%%%%%%%%%%%%%%%%%%
\begin{figure}[t]
\bc\includegraphics[width=9.5cm]{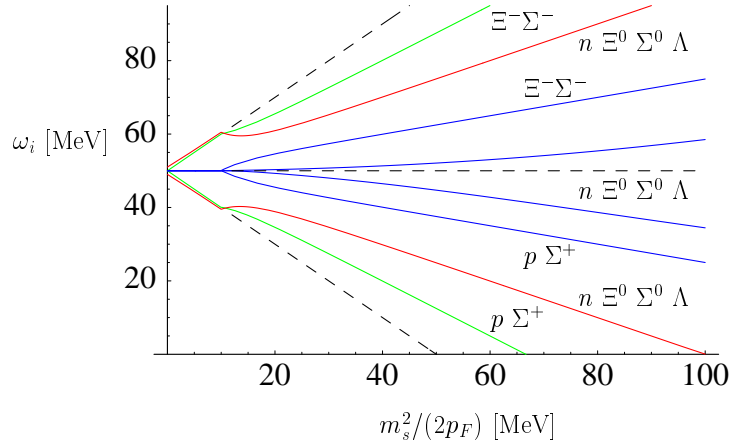}\ec
\caption{\label{fig_cfl_spec}
This figure shows the fermion spectrum in the CFL phase. For 
$m_s=0$ there are eight fermions with gap $\Delta$ and one
fermion with gap $2\Delta$ (not shown). Without kaon condensation
gapless fermion modes appear at $\mu_s=\Delta$ (dashed lines).
With kaon condensation gapless modes appear at $\mu_s=4\Delta/3$.}
\end{figure}
%%%%%%%%%%%%%%%%%%%%%%%%%%%%%%%%%%%%%%%%%%%%%%%%%%%%%%%%%%%%%%%%%%%%

%%%%%%%%%%%%%%%%%%%%%%%%%%%%%%%%%%%%%%%%%%%%%%%%%%%%%%%%%%%%%%%%%%
\subsection{Fermions in the CFL phase}
\label{sec_gCFL}
%%%%%%%%%%%%%%%%%%%%%%%%%%%%%%%%%%%%%%%%%%%%%%%%%%%%%%%%%%%%%%%%%%

 So far we have only studied Goldstone modes in the CFL phase.
However, as the strange quark mass is increased it is possible
that some of the fermion modes become light or even gapless
\cite{Alford:2003fq}. In order to study this question we 
have to include fermions in the effective field theory. 
The effective lagrangian for fermions in the CFL phase
is \cite{Kryjevski:2004jw,Kryjevski:2004kt}
\bea 
\label{l_bar}
{\mathcal L} &=&  
 {\rm Tr}\left(N^\dagger iv^\mu D_\mu N\right) 
 - D{\rm Tr} \left(N^\dagger v^\mu\gamma_5 
               \left\{ {\mathcal A}_\mu,N\right\}\right)
 - F{\rm Tr} \left(N^\dagger v^\mu\gamma_5 
               \left[ {\mathcal A}_\mu,N\right]\right)
  \nonumber \\
 & &  \mbox{} + \frac{\Delta}{2} \left\{ 
     \left( {\rm Tr}\left(N_LN_L \right) 
   - \left[ {\rm Tr}\left(N_L\right)\right]^2 \right)  
   - (L\leftrightarrow R) + h.c.  \right\}.
\eea
$N_{L,R}$ are left and right handed baryon fields in the 
adjoint representation of flavor $SU(3)$. The baryon fields 
originate from quark-hadron complementarity \cite{Schafer:1998ef}. 
We can think of $N$ as describing a quark which is surrounded 
by a diquark cloud, $N_L \sim q_L\langle q_L q_L\rangle$. The 
covariant derivative of the nucleon field is given by $D_\mu N
=\partial_\mu N +i[{\mathcal V}_\mu,N]$. The vector and axial-vector 
currents are 
\be
\label{v-av}
 {\mathcal V}_\mu = -\frac{i}{2}\left\{ 
  \xi \partial_\mu\xi^\dagger +  \xi^\dagger \partial_\mu \xi 
  \right\}, \hspace{1cm}
{\mathcal A}_\mu = -\frac{i}{2} \xi\left(\nabla_\mu 
    \Sigma^\dagger\right) \xi , 
\ee
where $\xi$ is defined by $\xi^2=\Sigma$. It follows that $\xi$ 
transforms as $\xi\to L\xi U(x)^\dagger=U(x)\xi R^\dagger$ with 
$U(x)\in SU(3)_V$. For pure $SU(3)$ flavor transformations $L=R=V$ 
we have $U(x)=V$. $F$ and $D$ are low energy constants that 
determine the baryon axial coupling. In perturbative QCD we
find $D=F=1/2$.

 The effective theory given in equ.~(\ref{l_bar}) can be derived 
from QCD in the weak coupling limit. However, the structure of the 
theory is completely determined by chiral symmetry, even if the 
coupling is not weak. In particular, there are no free parameters
in the baryon coupling to the vector current. Mass terms are also
strongly constrained by chiral symmetry. The effective chemical 
potentials $(X_L,X_R)$ appear as left and right-handed gauge 
potentials in the covariant derivative of the nucleon field.
We have 
\bea
\label{V_X}
 D_0N     &=& \partial_0 N+i[\Gamma_0,N], \\
 \Gamma_0 &=& -\frac{i}{2}\left\{ 
  \xi \left(\partial_0+ iX_R\right)\xi^\dagger + 
  \xi^\dagger \left(\partial_0+iX_L\right) \xi 
  \right\}, \nonumber 
\eea
where $X_L=MM^\dagger/(2p_F)$ and $X_R=M^\dagger M/(2p_F)$ as before.
$(X_L,X_R)$ covariant derivatives also appears in the axial vector 
current given in equ.~(\ref{v-av}).

 We can now study how the fermion spectrum depends on the quark mass.
In the CFL state we have $\xi=1$. For $\mu_s=0$ the baryon octet
has an energy gap $\Delta$ and the singlet has gap $2\Delta$. As a 
function of $\mu_s$ the excitation energy of the proton and neutron 
is lowered, $\omega_{p,n}=\Delta-\mu_s$, while the energy of the 
cascade states $\Xi^-,\Xi^0$ particles is raised, $\omega_{\Xi}=
\Delta+\mu_s$. All other excitation energies are independent of 
$\mu_s$. As a consequence we find gapless $(p,n)$ and $(\Xi^-,
\Xi^0)^{-1}$ excitations at $\mu_s=\Delta$. 

  The situation is more complicated when kaon condensation is taken 
into account. In the kaon condensed phase there is mixing in the 
$(p,\Sigma^+,\Sigma^-,\Xi^-)$ and $(n,\Sigma^0,\Xi^0,\Lambda^8,
\Lambda^0)$ sector. For $m_K^0\ll\mu_s\ll \Delta$ the spectrum is 
given by
\be
\omega_{p\Sigma^\pm\Xi^-}= \left\{
 \begin{array}{c}
 \Delta \pm \frac{3}{4}\mu_s, \\
 \Delta \pm \frac{1}{4}\mu_s,
\end{array}\right.  \hspace{0.75cm}
\omega_{n\Sigma^0\Xi^0\Lambda} = \left\{
 \begin{array}{c}
   \Delta \pm \frac{1}{2}\mu_s ,\\ 
   \Delta , \\
   2\Delta .
 \end{array} \right. 
\ee
Numerical results for the eigenvalues are shown in Fig.~\ref{fig_cfl_spec}. 
We observe that mixing within the charged and neutral baryon sectors leads 
to level repulsion. There are two modes that become light in the CFL window 
$\mu_s\leq 2\Delta$. One mode is a linear combination of proton and 
$\Sigma^+$ particles, as well as $\Xi^-$ and $\Sigma^-$ holes, and 
the other mode is a linear combination of the neutral baryons $(n,
\Sigma^0,\Xi^0,\Lambda^8,\Lambda^0)$.

%%%%%%%%%%%%%%%%%%%%%%%%%%%%%%%%%%%%%%%%%%%%%%%%%%%%%%%%%%%
\begin{figure}[t]
\bc\includegraphics[width=6.3cm]{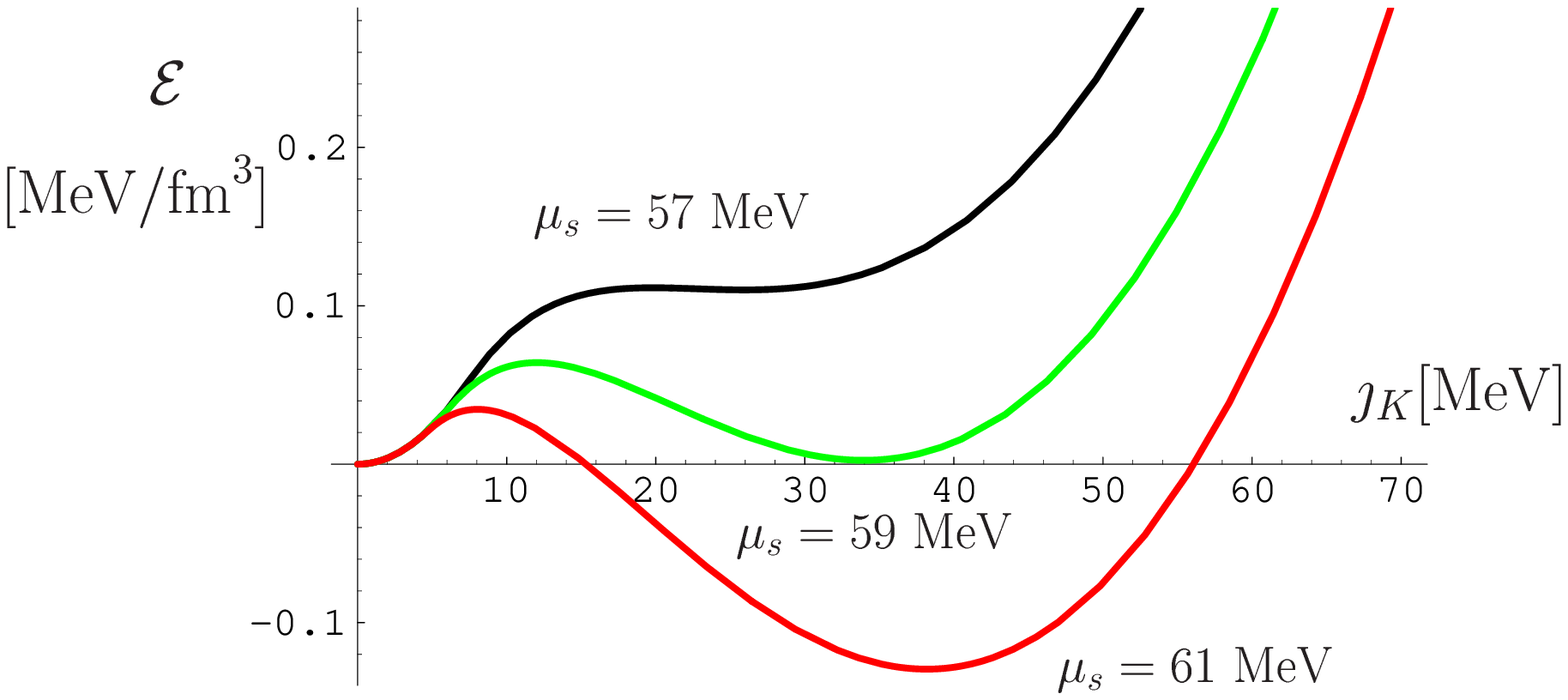}\hspace*{-0.3cm}
\includegraphics[width=6.3cm]{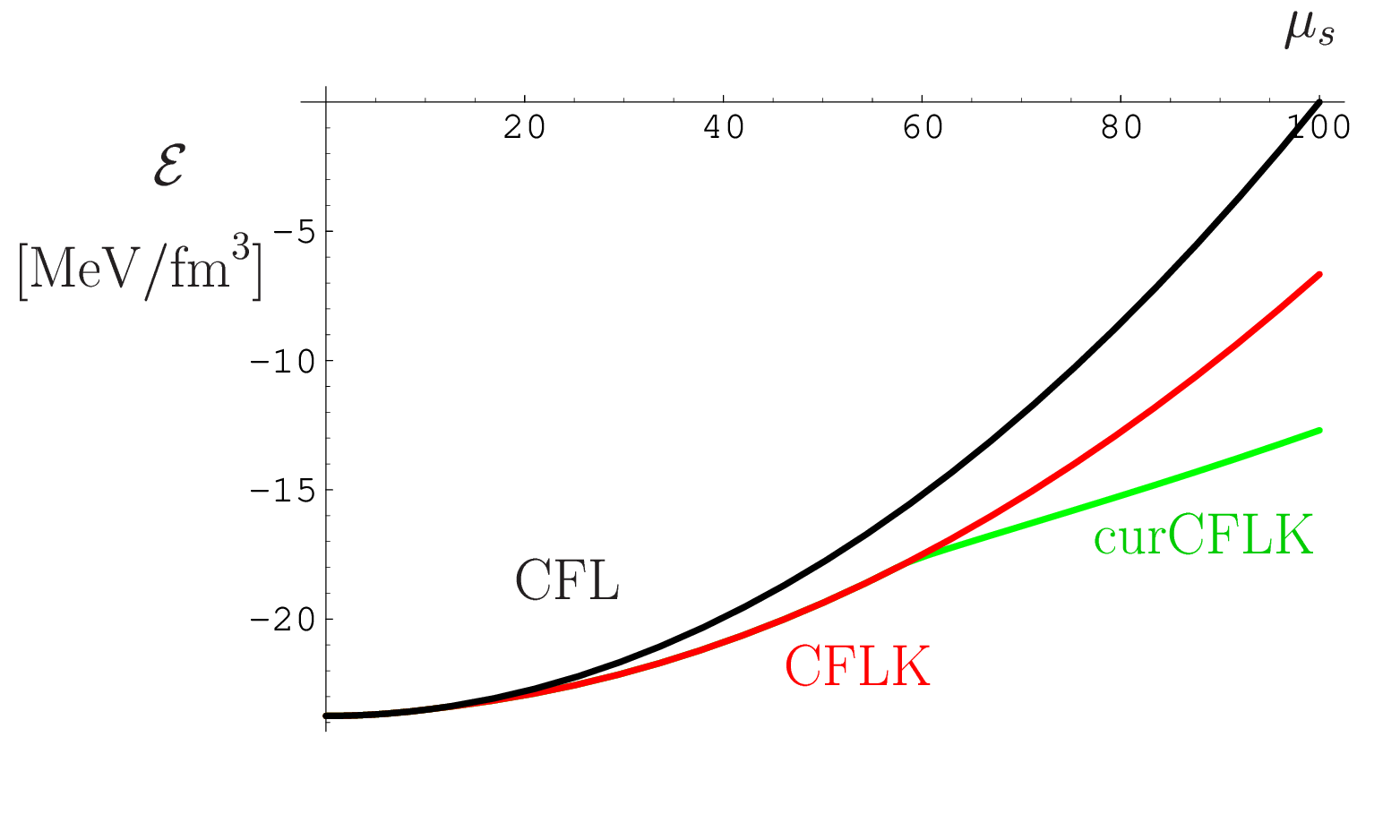}\ec
\caption{Left panel: Energy density as a function of the current 
$\jmath_K$ for several different values of $\mu_s=m_s^2/(2p_F)$ 
close to the phase transition. Right panel: Ground state energy 
density as a function of $\mu_s$. We show the CFL phase, the 
kaon condensed CFL (KCFL) phase, and the supercurrent state
(curKCFL).}
\label{fig_jfct}
\end{figure}
%%%%%%%%%%%%%%%%%%%%%%%%%%%%%%%%%%%%%%%%%%%%%%%%%%%%%%%%%%%

%%%%%%%%%%%%%%%%%%%%%%%%%%%%%%%%%%%%%%%%%%%%%%%%%%%%%%%%%%%%%%%%%
\subsection{Meson supercurrent state}
\label{sec_cur}
%%%%%%%%%%%%%%%%%%%%%%%%%%%%%%%%%%%%%%%%%%%%%%%%%%%%%%%%%%%%%%%%%

 Recently, several groups have shown that gapless fermion modes lead 
to instabilities in the current-current correlation function 
\cite{Huang:2004bg,Casalbuoni:2004tb}. Motivated by these results we 
have examined the stability of the kaon condensed phase against the 
formation of a non-zero current \cite{Schafer:2005ym,Kryjevski:2005qq}.
Consider a spatially varying $U(1)_Y$ rotation of the maximal kaon 
condensate
\be 
U(x)\xi_{K^0} U^\dagger (x) = \left(
 \begin{array}{ccc}
 1 & 0 & 0 \\
 0 & 1/\sqrt{2} & ie^{i\phi_K(x)}/\sqrt{2} \\
 0 & ie^{-i\phi_K(x)}/\sqrt{2} & 1/\sqrt{2} 
\end{array} \right).
\ee
This state is characterized by non-zero currents
\be
\label{cur}
\vec{\cal V} =  \frac{1}{2}\left(\vec{\nabla} \phi_K\right) \left(
 \begin{array}{ccc}
0 & 0 & 0 \\
0 & 1 & 0 \\
0 & 0 & -1 
\end{array} \right),\hspace{0.3cm}
\vec{\cal A} =  \frac{1}{2}\left(\vec{\nabla} \phi_K\right) \left(
 \begin{array}{ccc}
0 & 0 & 0 \\
0 & 0 & -ie^{i\phi_K} \\
0 & ie^{-i\phi_K} & 0 
\end{array} \right).
\ee
In the following we compute the vacuum energy as a function 
of the kaon current $\vec{\jmath}_K=\vec\nabla\phi_K$. The meson 
part of the effective lagrangian gives a positive contribution
\be
{\cal E}=\frac{1}{2}v_\pi^2f_\pi^2\jmath_K^2 .
\ee
A negative contribution can arise from gapless fermions. In order 
to determine this contribution we have to calculate the fermion spectrum 
in the presence of a non-zero current. The relevant part of the
effective lagrangian is  
\bea 
{\cal L} &=& {\rm Tr}\left(N^\dagger iv^\mu D_\mu N\right)
 + {\rm Tr}\left(N^\dagger \gamma_5 \left( \rho_A+\vec{v}\cdot
      \vec{\cal A}\right) N\right) \nonumber \\
 & & \mbox{}  
 +\frac{\Delta}{2} \left\{ {\rm Tr}\left(N N\right) -
  {\rm Tr}\left(N\right){\rm Tr}\left(N\right)+ h.c.\right\},
\eea
where we have used $D=F=1/2$. The covariant derivative is 
$D_0N=\partial_0N+i[\rho_V,N]$ and $D_iN=\partial_i N +i
\vec{v}\cdot[\vec{\cal V},N]$ with $\vec{\cal V},\vec{\cal A}$ given in 
equ.~(\ref{cur}) and  
\be 
\rho_{V,A} = \frac{1}{2}\left\{ 
  \xi \frac{M^\dagger M}{2p_F}\xi^\dagger \pm 
  \xi^\dagger \frac{MM^\dagger}{2p_F} \xi 
  \right\}.
\ee
The vector potential $\rho_V$ and the vector current $\vec{\cal V}$
are diagonal in flavor space while the axial potential $\rho_A$ 
and the axial current $\vec{\cal A}$ lead to mixing. The fermion 
spectrum is quite complicated. The dispersion relation of the lowest 
mode is approximately given by
\be
\label{disp_ax}
\omega_l = \Delta +\frac{(l-l_0)^2}{2\Delta}-\frac{3}{4}
  \mu_s -\frac{1}{4}\vec{v}\cdot\vec{\jmath}_K,
\ee
where $l=\vec{v}\cdot\vec{p}-p_F$ and we have expanded $\omega_l$ 
near its minimum $l_0=(\mu_s+\vec{v}\cdot\vec{\jmath}_K)/4$.
Equation (\ref{disp_ax}) shows that there is a gapless mode if 
$\mu_s>4\Delta/3-\jmath_K/3 $. The contribution of the gapless mode 
to the vacuum energy is 
\be
\label{e_fct} 
{\cal E} = \frac{\mu^2}{\pi^2}\int dl \int 
 \frac{d\Omega}{4\pi} \;\omega_l \theta(-\omega_l) ,
\ee
where $d\Omega$ is an integral over the Fermi surface. The integral 
in equ.~(\ref{e_fct}) receives contributions from one of the pole caps 
on the Fermi surface. The result has exactly the same structure as the 
energy functional of a non-relativistic two-component Fermi liquid with 
non-zero polarization, see Sect.~\ref{sec_atom}. Introducing dimensionless 
variables 
\be 
\label{x+h}
 x = \frac{\jmath_K}{a\Delta}, \hspace{0.5cm}
 h = \frac{3\mu_s-4\Delta}{a\Delta}.
\ee
we can write ${\cal E} = c\, {\cal N} f_h(x)$ with
\be
\label{cur_fct}
 f_h(x) = x^2-\frac{1}{x}\left[
   (h+x)^{5/2}\Theta(h+x) - (h-x)^{5/2}\Theta(h-x) \right] .
\ee
We have defined the constants
\be
\label{consts}
 c = \frac{2}{15^4c_\pi^3 v_\pi^6},\hspace{0.45cm}
 {\cal N} =   \frac{\mu^2\Delta^2}{\pi^2},\hspace{0.45cm}
 a = \frac{2}{15^2 c_\pi^2 v_\pi^4}, 
\ee
where $c_\pi = (21-8\log(2))/36$ is the numerical coefficient 
that appears in the weak coupling result for $f_\pi$. 
According to the analysis in \cite{Son:2005qx} the function
$f_h(x)$ develops a non-trivial minimum if $h_1<h<h_2$ with 
$h_1\simeq -0.067$ and $h_2\simeq 0.502$. In perturbation 
theory we find $a=0.43$ and the kaon condensed ground state
becomes unstable for $(\Delta- 3\mu_s/4) < 0.007\Delta$. 

%%%%%%%%%%%%%%%%%%%%%%%%%%%%%%%%%%%%%%%%%%%%%%%%%%%%%%%%%%%
\begin{figure}[t]
\bc\includegraphics[width=8.0cm]{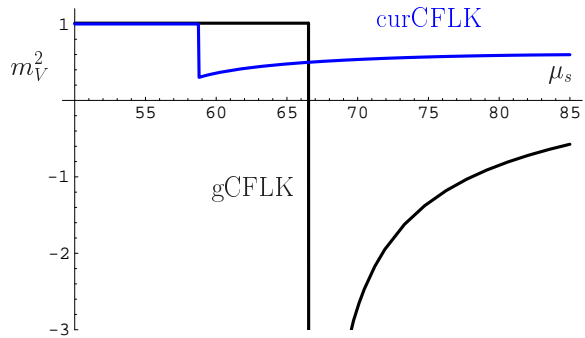}\ec
\caption{
Screening mass of flavor gauge fields in the CFL phase.
The two curves show the second derivative of the effective 
potential with respect to the current at the origin and at 
the minimum of the potential. }
\label{fig_kstar}
\end{figure}
%%%%%%%%%%%%%%%%%%%%%%%%%%%%%%%%%%%%%%%%%%%%%%%%%%%%%%%%%%%

 The energy density as a function of the current and the 
groundstate energy density as a function of $\mu_s$ are 
shown in Fig.~\ref{fig_jfct}. In these plots we have included 
the contribution of a baryon current $\jmath_B$, as suggested 
in \cite{Kryjevski:2005qq}. In this case we have to minimize the 
energy with respect to two currents. The solution is of the form
$\jmath_B\sim \jmath_K$. The figure shows the dependence on 
$\jmath_K$ for the optimum value of $\jmath_B$. We have not properly 
implemented electric charge neutrality. Since the gapless mode is 
charged, enforcing electric neutrality will significantly suppress 
the magnitude of the current \cite{Kryjevski:2006}. We have also 
not included the possibility that the neutral mode becomes gapless. 
This will happen at somewhat larger values of $\mu_s$.

 We note that the ground state has no net current. This is clear 
from the fact that the ground state satisfies $\delta {\cal E}/\delta 
(\vec\nabla\phi_K)=0$. As a consequence the meson current is canceled 
by an equal but opposite contribution from gapless fermions. We also 
expect that the ground state has no chromomagnetic instabilities. 
From the effective lagrangian we can compute the screening length
of a $SU(3)_F$ gauge field. In the CFL phase the isospin and hypercharge 
screening masses are 
\be
m_V^2=\left. \frac{\partial^2\mathcal E}{\partial \jmath_K^2}
  \right|_{\jmath_K =0} 
 = v_\pi^2 f_\pi^2\left(1 - \frac{5}{8\sqrt{h}}\theta(h)\right),
\ee
which shows the magnetic instability for $h>0$ and has the characteristic 
square root singularity observed in microscopic calculations 
\cite{Huang:2004bg}. In Fig.~\ref{fig_kstar} we show the screening 
mass in the kaon condensed CFL phase and the supercurrent phase
as a function of $\mu_s$. We observe that there is an instability 
in the homogeneous phase, but the instability disappears in the 
supercurrent state.

%%%%%%%%%%%%%%%%%%%%%%%%%%%%%%%%%%%%%%%%%%%%%%%%%%%%%%%%%%%%%%%%%
\subsection{Instanton effects}
\label{sec_inst}
%%%%%%%%%%%%%%%%%%%%%%%%%%%%%%%%%%%%%%%%%%%%%%%%%%%%%%%%%%%%%%%%%
  
 The results discussed in Sects.~\ref{sec_kcond}-\ref{sec_cur} are 
based on an effective chiral theory of the CFL phase. The coefficients
of the effective theory are determined by matching to perturbative
QCD calculations. We find that the chiral coefficients are ``natural'',
i.e.~their numerical value agrees with simple dimensional estimates. 
For example, $f_\pi^2$ is given, up to a coefficient of order one, 
by the density of states on the Fermi surface. This suggest that 
many of our results are valid even if the QCD coupling is not weak. 

\begin{figure}[t]
\bc\includegraphics[width=8cm]{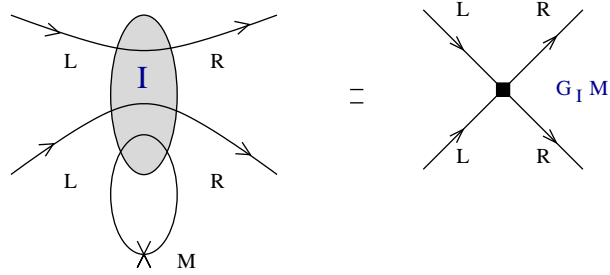}\ec
\caption{\label{fig_ibcs}
Instanton contribution to the BCS four-fermion operator 
in QCD with three flavors.}
\end{figure}

 A possible exception is the linear mass term in the effective chiral
theory. The coefficient $B$ in equ.~(\ref{l_cheft}) vanishes to all 
orders in perturbation theory, but $B$ receives non-perturbative 
contributions from instantons. In QCD with three flavors instantons 
induce a four-fermion operator \cite{Schafer:2002ty}.
\bea
\label{l_nf2}
{\cal L} &=& \int n(\rho,\mu)d\rho\, 
   \frac{2(2\pi\rho)^4\rho^3}{4(N_c^2-1)}
 \epsilon_{f_1f_2f_3}\epsilon_{g_1g_2g_3}M_{f_3 g_3}
 \left( \frac{2N_c-1}{2N_c}
  (\bar\psi_{R,f_1} \psi_{L,g_1})
  (\bar\psi_{R,f_2} \psi_{L,g_2}) \right. \nonumber \\
& & \hspace{1cm}\mbox{}\left. - \frac{1}{8N_c}
  (\bar\psi_{R,f_1} \sigma_{\mu\nu} \psi_{L,g_1})
  (\bar\psi_{R,f_2} \sigma_{\mu\nu} \psi_{L,g_2})
  + (M\leftrightarrow M^\dagger, 
     L \leftrightarrow R ) \right)  ,
\eea 
see Fig.~\ref{fig_ibcs}. Here, $f_i,g_i$ are flavor indices and 
$\sigma_{\mu\nu}=\frac{i}{2}[\gamma_\mu,\gamma_\nu]$. The instanton size
distribution $n(\rho,\mu)$ is given by
%%%%%%%%%%%%%%%%%%%%%%%%%%%%%%%%%%%%%%%%%%%%%%%%%%%%%%%%%%%%%%%%%%%%
%%%%%%%%%%%%%%%%%%%%%%%%%%%%%%%%%%%%%%%%%%%%%%%%%%%%%%%%%%%%%%%%%%%%
\bea
\label{G_I}
  n(\rho,\mu) &=& C_{N} \ \left(\frac{8\pi^2}{g^2}\right)^{2N_c} 
 \rho^{-5}\exp\left[-\frac{8\pi^2}{g(\rho)^2}\right]
 \exp\left[-N_f\rho^2\mu^2\right],\\
 && C_{N} \;=\; \frac{0.466\exp(-1.679N_c)1.34^{N_f}}
    {(N_c-1)!(N_c-2)!}\, . 
\eea
At zero density, the $\rho$ integral in equ.~(\ref{l_nf2}) diverges
for large $\rho$. This is the well-known infrared problem of the 
semi-classical approximation in QCD. At large chemical potential, 
however, large instantons are suppressed and the typical instanton size 
is $\rho\sim \mu^{-1} \ll \Lambda^{-1}_{QCD}$ where $\Lambda_{QCD}$ is
the QCD scale parameter. The instanton contribution to the vacuum energy 
density is
\bea
\label{E_I}
{\cal E} &=& -\int n(\rho,\mu) d\rho\,
 \frac{16}{3}(\pi\rho)^4\rho^3 
 \left[\frac{3\sqrt{2}\pi}{g}\Delta
     \left(\frac{\mu^2}{2\pi^2}\right)\right]^2
  {\rm Tr}\left[M+M^\dagger\right] .
\eea
This result can be matched against the $O(M)$ term in the effective 
chiral lagrangian. We find
\be 
\label{B}
 B = C_N\frac{8\pi^4}{3}\frac{\Gamma(6)}{3^6}
 \left[\frac{3\sqrt{2}\pi}{g}\Delta
     \left(\frac{\mu^2}{2\pi^2}\right)\right]^2
  \left(\frac{8\pi^2}{g^2}\right)^{6}
  \left(\frac{\Lambda_{QCD}}{\mu}\right)^{12}\Lambda_{QCD}^{-3},
\ee
where we have performed the integral over the instanton
size $\rho$ using the one-loop beta function. The coefficient
$B$ is related to the quark-anti-quark condensate, $\langle
\bar{\psi}\psi\rangle =-2B$. The linear mass terms contributes
to the mass of the kaon, $\delta m_K^2 \sim (4B/f_\pi^2)mm_s$,
and tends to inhibit kaon condensation. At moderate density 
this contribution is quite uncertain because the result is 
very sensitive to the value of the strong coupling constant. 
Using the one-loop running coupling and $\Lambda_{QCD}\simeq
200$ MeV gives values as large as $\delta m_K \sim$ 100 MeV. 
This is almost certainly an overestimate because the main 
contribution comes from large instantons with size $\rho\sim 
0.5$ fm. If the average instanton size is constrained to be 
less than the phenomenological value at zero density, $\bar{\rho}
=0.35$ fm, then we find $\delta m_K < 10$ MeV. 

% The instanton term supresses kaon condensation if $2Bm_s>\mu_s^2
%f_\pi^2$. For this consition to be satisfied in the whole CFL window 
%$\mu_s<2\Delta$ requires $2Bm_s > 4f_\pi^2\Delta^2$. This implies
%that the instanton contribution to the energy density is comparable 
%to the CFL condensation energy.

%%%%%%%%%%%%%%%%%%%%%%%%%%%%%%%%%%%%%%%%%%%%%%%%%%%%%%%%%%%%%%%%%
\subsection{Microscopic models}
\label{sec_mic}
%%%%%%%%%%%%%%%%%%%%%%%%%%%%%%%%%%%%%%%%%%%%%%%%%%%%%%%%%%%%%%%%%

 In this section we shall compare some of our results with 
microscopic calculations based on Nambu-Jona-Lasinio (NJL)
models, see for example \cite{Alford:2003fq,Ruster:2004eg}.
A natural choice is to model the interaction between quarks
by a local four-fermion interaction with the quantum numbers
of one-gluon exchange
\be
\label{njl}
 {\mathcal L} = \bar\psi \left( i\dslash +\mu\gamma_0 - M\right)\psi
  + G \left(\bar\psi \gamma_\mu\lambda^a\psi\right)^2.
\ee
This lagrangian has the symmetries of QCD except that the $SU(3)$ 
color symmetry is global rather than local, and the axial $U(1)_A$
is not anomalous. The effects of the anomaly can be taken into 
account by adding the instanton induced four-fermion interaction
given in equ.~(\ref{l_nf2}). Since color is a global symmetry the 
NJL model does not exhibit a Higgs mechanism and color superconductivity
leads to an extra octet of colored Goldstone bosons. 

%%%%%%%%%%%%%%%%%%%%%%%%%%%%%%%%%%%%%%%%%%%%%%%%%%%%%%%%%%%%%%%%%
\begin{table}[t]
\tbl{Effective chemical potential for the different quark modes.
The quark modes are labeled by their color (rgb) and flavor (uds).
We also show the $\tilde{Q}$ charge and the corresponding baryon 
mode. The leading order result in the neutral CFL phase corresponds
to $\mu_e=\mu_3=0$ and $\mu_8=-\mu_s$.}
{\begin{tabular}{|c|r|l|l|l|}\hline 
mode & $\tilde{Q}$ & effective chemical potential 
     &  leading order  & baryon mode \\ \hline \hline 
 {ru}& 0 & $-\frac{2}{3}\mu_e+\frac{1}{2}\mu_3+\frac{1}{3}\mu_8$
     &     $\mu_0$ &  \\[1ex]
 {gd}& 0 & $+\frac{1}{3}\mu_e-\frac{1}{2}\mu_3+\frac{1}{3}\mu_8$
     &     $\mu_0$ &  $(\Lambda_0,\Lambda_8,\Sigma_0)$\\[1ex]
 {bs}& 0 & $+\frac{1}{3}\mu_e-\frac{2}{3}\mu_8-\mu_s$
     &     $\mu_0$ &  \\[3ex]
 {rd}&-1 & $+\frac{1}{3}\mu_e+\frac{1}{2}\mu_3+\frac{1}{3}\mu_8$
     &     $\mu_0$ &   \hspace{0.55cm}$\Sigma^-$         \\[1ex]
 {gu}& 1 & $-\frac{2}{3}\mu_e-\frac{1}{2}\mu_3+\frac{1}{3}\mu_8$
     &     $\mu_0$ &   \hspace{0.55cm}$\Sigma^+$         \\[3ex]  
 {rs}&-1 & $+\frac{1}{3}\mu_e+\frac{1}{2}\mu_3+\frac{1}{3}\mu_8-\mu_s$
     &     $\mu_0-\mu_s$ &   \hspace{0.55cm}$\Xi^-$                   \\[1ex]
 {bu}& 1 & $-\frac{2}{3}\mu_e-\frac{2}{3}\mu_8$
     &     $\mu_0+\mu_s$ &   \hspace{0.55cm}$p$ \\[3ex]
 {gs}& 0 & $+\frac{1}{3}\mu_e-\frac{1}{2}\mu_3+\frac{1}{3}\mu_8-\mu_s$
     &     $\mu_0-\mu_s$ &   \hspace{0.55cm}$\Xi^0$ \\[1ex]
 {bd}& 0 & $+\frac{1}{3}\mu_e-\frac{2}{3}\mu_8$
     &     $\mu_0+\mu_s$ &   \hspace{0.55cm}$n$ \\ \hline 
\end{tabular}
\label{table1}}
\end{table}
%%%%%%%%%%%%%%%%%%%%%%%%%%%%%%%%%%%%%%%%%%%%%%%%%%%%%%%%%%%%%%%%%

 The pairing ansatz is usually taken to be 
\be
\label{cfl_123}
\langle \psi^a_i C\gamma_5 \psi^b_j \rangle \sim 
 \Delta_1 \epsilon^{ab1}\epsilon_{ij1} 
+ \Delta_2 \epsilon^{ab2}\epsilon_{ij2}  
+ \Delta_3 \epsilon^{ab3}\epsilon_{ij3}.
\ee
This ansatz reduces to the CFL ansatz for $\Delta_1=\Delta_2=\Delta_3$
and allows for the breaking of hypercharge and isospin. The ansatz
is not sufficiently rich to allow for kaon condensation. The chiral 
field $\Sigma$ can be defined as 
\be 
 \Sigma = XY^\dagger ,
\ee
where $X,Y$ are related to the left and right handed condensates
\bea 
\langle (\psi_L)^a_i C (\psi_L)^b_j \rangle \epsilon^{abc}\epsilon_{ijk}
  &\sim&  X_k^c ,\\
\langle (\psi_R)^a_i C (\psi_R)^b_j \rangle \epsilon^{abc}\epsilon_{ijk}
  &\sim&  Y_k^c .
\eea
In order to study kaon condensation we have to allow the left and right 
handed order parameters to be independent. This problem was studied in 
two papers by Buballa and Forbes \cite{Buballa:2004sx,Forbes:2004ww}.

 Color and electric charge neutrality are enforced by adding chemical 
potentials $(\mu_3,\mu_8,\mu_e)$ for color isospin, color hypercharge 
and electric charge. In most microscopic calculations the effect of 
the strange quark mass is taken into account by considering an effective 
chemical potential $\mu_s=m_s^2/(2p_F)$ for the strange quark. In 
Table \ref{table1} we list the chemical potentials for the different 
quark states. The quarks are labeled by color $(rgb)$ and flavor 
$(uds)$. They are grouped into the main pairing channels in the CFL 
phase. We also show the leading order result in the charge neutral 
CFL phase. In this case we find $\mu_e=\mu_3=0$ and $\mu_8=-\mu_s$ 
and the dominant pair breaking stress occurs in the $(rs)-(bu)$ and 
$(gs)-(bd)$ sector. 

 We can associate the quark states in the microscopic theory with 
the baryon fields in the effective theory by computing their 
quantum numbers under the unbroken $SU(3)_F$ symmetry. The effective 
theory is formulated in terms of gauge invariant fields and does not
require color chemical potentials. In weak coupling we can derive 
the effective theory by integrating out gluonic degrees of freedom. 
In this case the equations of motion for the color gauge potential
automatically enforce color neutrality.  

 We observe that the leading order results in the NJL calculation 
agree with the results derived from the effective field theory. 
The comparison was extended to the kaon condensed phase by Forbes
\cite{Forbes:2004ww}. He finds that the $bu$ (proton) and $bd$
(neutron) states split, and that the pair breaking stress on both 
states is reduced. The critical $\mu_s$ for gapless states is 
$\mu_s=1.2\Delta$, in qualitative agreement with the leading 
order EFT result $\mu_s=4\Delta/3$. 

 Microscopic calculations show that in the CFL phase the gap parameters 
$\Delta_{1,2,3}$ are very similar even if the pair breaking stress 
$\mu_s$ is close to the value of the gap \cite{Alford:2003fq}. The
main effect is a reduction of the average gap which is presumably 
related to the decrease in the common Fermi momentum that occurs
if $\mu_s$ is increased at constant $\mu_B$. In the gapless CFL phase, 
on the other hand, the splitting between the different gap parameters 
becomes significant. Neither one of these two effects is included in 
the EFT calculations discussed in Sects.~\ref{sec_kcond}-\ref{sec_cur}.
The response of the gap to the trace part of $MM^\dagger/p_F^2$ can
be taken into account by adding a higher order operator to the 
effective lagrangian equ.~(\ref{l_bar}), but the coefficient of this
term is model dependent. Indeed, in perturbative QCD the gap will
likely increase if $p_F$ is lowered. The splitting between the gaps
in the gapless CFL is due to the fact that gapless regions on the 
Fermi surface no longer contribute to pairing. In the EFT treatment
this effect is not included because the current is assumed to be 
much smaller than the gap. Clearly, in the regime where the number 
of gapless states is large the back-reaction on the gap has to be 
taken into account.

%%%%%%%%%%%%%%%%%%%%%%%%%%%%%%%%%%%%%%%%%%%%%%%%%%%%%%%%%%%%%%%%%
\section{Cold atomic systems}
\label{sec_atom}
%%%%%%%%%%%%%%%%%%%%%%%%%%%%%%%%%%%%%%%%%%%%%%%%%%%%%%%%%%%%%%%%%

 A nice system in which stressed pairing can be studied in the 
laboratory is a cold dilute gas of fermionic atoms. Using Feshbach
resonances it is possible to tune the interaction between the 
Bose-Einstein (BEC) limit of tightly bound diatomic molecules
and the Bardeen-Cooper-Schrieffer (BCS) limit of weakly correlated
Cooper pairs. The most interesting part of the phase diagram is 
the crossover regime where the two-body scattering length diverges. 

\begin{figure}[t]
\bc\includegraphics[width=7.5cm]{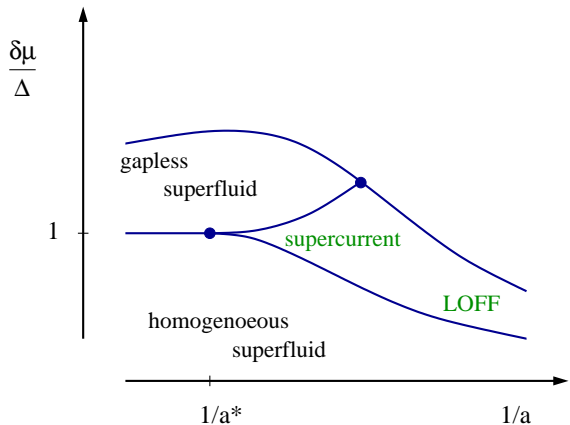}\ec
\caption{\label{fig_ph}
Conjectured phase diagram for a polarized cold atomic Fermi gas
as a function of the scattering length $a$ and the difference 
in the chemical potentials $\delta\mu=\mu_\uparrow-\mu_\downarrow$, 
from Son \& Stephanov (2005).}
\end{figure}

 A conjectured (and, most likely, oversimplified) phase diagram 
for a polarized gas is shown in Fig.~\ref{fig_ph}. In the BEC 
limit the gas consists of tightly bound spin singlet molecules. 
Adding an extra up or down spin requires energy $\Delta$. For 
$|\delta\mu|=|\mu_\uparrow-\mu_\downarrow|>\Delta$ the system 
is a homogeneous mixture of a Bose condensate and a fully polarized
Fermi gas. One can show that in the dilute limit this mixture is 
stable with regard to phase separation \cite{Viverit:2000}.

%%%%%%%%%%%%%%%%%%%%%%%%%%%%%%%%%%%%%%%%%%%%%%%%%%%%%%%%%%%%%%%%%%%%
%%%%%%%%%%%%%%%%%%%%%%%%%%%%%%%%%%%%%%%%%%%%%%%%%%%%%%%%%%%%%%%%%%%%

 The Bose-Fermi-gas mixture is a gapless superfluid. We can ask, 
along the lines of Sect.~\ref{sec_cur}, whether the system is 
stable with respect to the formation of a non-zero supercurrent. 
We can study this question using the effective lagrangian
\be
\label{leff_gbcs}
  {\cal L} =  \psi^\dagger \Big(i\partial_0 - \epsilon(-i\vec{\partial}) 
  - i(\vec{\partial}\varphi)\cdot
   \frac{\stackrel{\scriptstyle\leftrightarrow}{\partial}}{2m} 
    \Big)\psi
  + \frac{f_t^2}{2} \dot\varphi^2 - \frac{f^2}{2} (\vec{\partial}\varphi)^2.
\ee
Here, $\psi$ describes a gapless fermion with dispersion law $\epsilon
(\vec{p})$ and $\varphi$ is the superfluid Goldstone mode. The low energy 
parameters $f_t$ and $f$ are related to the density and the velocity of 
sound. Similar to the chiral theory discussed in Sect.~\ref{sec_cur}
the p-wave coupling of the fermions to the Goldstone boson is governed
by the $U(1)$ symmetry of the theory. 

 Setting up a current $\vec{v}_s=\vec{\partial}\varphi/m$ requires energy
$f^2m^2v_s^2/2$. The contribution from fermions can be computed using 
the fermion dispersion law in the presence of a non-zero current
\be
  \epsilon_v(\vec{p}) = \epsilon(\vec{p})
          + \vec{v}_s\cdot\vec{p}  -\delta\mu\, .
\ee
The total free energy is 
\be 
F(v_s) = \frac{1}{2} n m v_s^2 + 
  \int \frac{d^3p}{(2\pi)^3} \, \epsilon_v(\vec{p}) 
    \Theta\left(-\epsilon_v(\vec{p})\right),
\ee
where $n$ is the density and we have used $f^2=n/m$. Son and Stephanov 
noticed that the stability of the gapless phase depends crucially on 
the nature of the dispersion law $\epsilon(p)$. For small momenta we 
can write $\epsilon(p)\simeq \epsilon_0+\alpha p^2+\beta p^4$. In the 
BEC limit $\alpha>0$ and the minimum of the dispersion curve is at 
$p=0$ while in the BCS limit $\alpha<0$ and the minimum is at $p\neq 0$.
In the latter case the density of states on the Fermi surface is
finite and the system is unstable with respect to the formation 
of a non-zero current. The free energy functional is of exactly 
the same type as the one given in equ.~(\ref{cur_fct}). On the 
other hand, if the minimum of the dispersion curve is at zero, 
then the density of states vanishes and there is no instability. 
As a consequence there is a critical point along the BEC-BCS 
line at which the instability will set in. 

%%%%%%%%%%%%%%%%%%%%%%%%%%%%%%%%%%%%%%%%%%%%%%%%%%%%%%%%%%%%%%%%%%%%
\begin{figure}[t]
\bc\includegraphics[width=9.5cm]{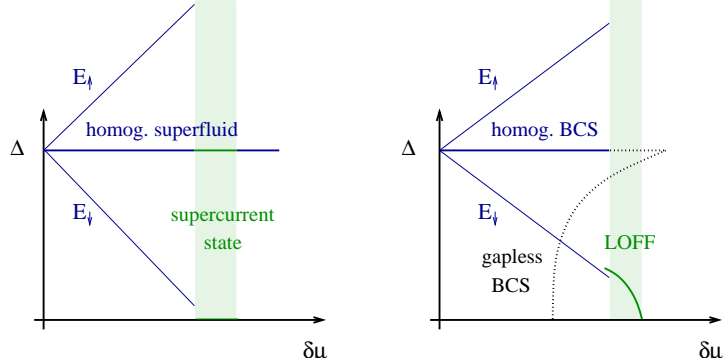}\ec
\caption{\label{fig_loff}
Schematic behavior of the gap parameter and the quasi-particle 
energies near the onset of the supercurrent state (left panel)
and in the BCS limit (right panel).}
\end{figure}
%%%%%%%%%%%%%%%%%%%%%%%%%%%%%%%%%%%%%%%%%%%%%%%%%%%%%%%%%%%%%%%%%%%%

 We can also analyze the system in the BCS limit. This analysis
goes back to the work of Larkin, Ovchninikov, Fulde and Ferell
\cite{Larkin:1964,Fulde:1964}, see the review \cite{Casalbuoni:2003wh}.   
First consider homogeneous solutions to the BCS gap equation for 
$\delta\mu\neq 0$. In the regime $\delta\mu<\Delta_0$ where $\Delta_0=
\Delta(\delta\mu\!=\!0)$ the gap equation has a solution with gap 
parameter $\Delta=\Delta_0$. The free energy of this solution is 
\be 
F = -\frac{N}{4} \left(\Delta^2-2\delta\mu^2 \right),
\ee
where $N$ is the density of states on the Fermi surface. For 
$\delta\mu>\Delta_0/2$ there is a second solution with $\Delta=
(2\delta\mu\Delta_0-\Delta_0^2)^{1/2}$, but this solution is 
a local maximum of the free energy. There are no gapless modes
in the stable BCS phase, but the unstable BCS phase is gapless,
see Fig.~\ref{fig_loff}. 

 For $\Delta_0>\delta\mu>\Delta_0/\sqrt{2}$ the gap equation has 
a non-trivial solution, but the free energy is higher than the 
free energy of the normal phase and the solution is only meta-stable. 
LOFF~studied whether it is possible to find a stable solution in 
which the gap has a spatially varying phase
\be 
\label{loff}
 \Delta(\vec{x})= \Delta e^{2i\vec{q}\cdot\vec{x}}.
\ee
This solution exists in the LOFF window $\delta\mu_1<\delta\mu<
\delta\mu_2$ with $\delta\mu_1=\Delta_0/\sqrt{2}\simeq 0.71
\Delta_0$ and $\delta\mu_2\simeq 0.754\Delta_0$. The LOFF momentum 
$q$ depends on $\delta\mu$. Near $\delta\mu_2$ we have $qv_F
\simeq 1.2\delta\mu$. The gap $\Delta$ goes to zero near $\delta\mu_2$
and reaches $\Delta\simeq 0.25\Delta_0$ at $\delta\mu_1$.
 
 Clearly, the LOFF solution is of the same type as the supercurrent 
state. The $U(1)$ of baryon number is spontaneously broken and 
the phase of the condensate has a non-zero gradient. The difference
is that in the supercurrent state the current is much smaller 
than the gap, $v_F(\nabla\varphi)\ll \Delta$, while in the LOFF phase 
$v_F(\nabla\varphi)> \Delta$ (and $v_F(\nabla\varphi)\gg \Delta$
near $\delta\mu_2$). In the supercurrent state the Fermi surface
is mostly gapped but a small shell near one of the pole caps is
ungapped. In the weakly coupled LOFF state there are gapless 
excitations over most of the Fermi surface but pairing takes place
near two rings on the northern and southern hemisphere. 

 In the LOFF state it is energetically favorable to take linear 
superpositions of equ.~(\ref{loff}) with more than one plane 
wave. As a consequence, the LOFF state has a crystalline structure 
and there are planes on which the order parameter has a node. 
In the supercurrent state the current is much smaller than the 
gap and the formation of nodes is not favored. This implies that
there will be at least one phase transition (not shown in 
Fig.~\ref{fig_ph}) that separates the supercurrent state from 
the LOFF phase.

%%%%%%%%%%%%%%%%%%%%%%%%%%%%%%%%%%%%%%%%%%%%%%%%%%%%%%%%%%%%%%%%%
\section{Outlook}
\label{sec_sum}
%%%%%%%%%%%%%%%%%%%%%%%%%%%%%%%%%%%%%%%%%%%%%%%%%%%%%%%%%%%%%%%%%

 Does QCD with three flavors more closely resemble the supercurrent 
state or the LOFF state of the cold atomic system? In the CFL phase 
gapless excitations first appear in a regime where the paired state
is stable with respect to the normal phase or other homogeneous
phases. The resulting instability can be resolved by the formation
of a Goldstone boson current $\jmath<\Delta$. Whether or not 
the current is not only numerically but also parametrically 
small compared to the gap depends on certain details that need
to be studied more carefully. In the kaon condensed phase the 
lowest mode is charged and the current is suppressed by charge
neutrality. If the neutral mode becomes gapless, too, or if kaons 
are not condensed the current is no longer suppressed. Once the 
current becomes comparable to the gap it may be more appropriate 
to characterize the system as a LOFF state~\cite{Casalbuoni:2005zp}. 
In particular, it is possible that nodes in the condensate appear 
and quark matter turns into a crystal~\cite{Alford:2000ze}. 

 Ultimately we would like to obtain observational evidence for 
exotic phases of dense matter. Polarized atomic Fermi gases
have been created in the laboratory but so far neither the 
supercurrent state nor the LOFF state have been observed. In 
the case of quark matter we need to identify unique signatures 
of the possible phases that can be compared to observational 
evidence. Much work in this direction remains to be done. 
 
Acknowledgments: The work presented in this review was performed
in collaboration with P.~Bedaque, D.~Kaplan, A.~Kryjevski, and 
K.~Schwenzer. I would like to thank M.~Alford and A.~Sedrakian 
for Organizing the workshop on ``Pairing in fermionic systems:
Beyond the BCS theory'' at the INT (Seattle). This work is supported 
in part by the US Department of Energy grant DE-FG-88ER40388.

\end{document}